\documentstyle[12pt,aaspp4]{article}

\def\gtorder{\mathrel{\raise.3ex\hbox{$>$}\mkern-14mu
                \lower0.6ex\hbox{$\sim$}}}
\def\ltorder{\mathrel{\raise.3ex\hbox{$<$}\mkern-14mu
                \lower0.6ex\hbox{$\sim$}}}
\def\eg{ {\it e.g.~}}

\begin{document}

\title{A TULLY-FISHER RELATION FOR S0 GALAXIES}

\author{Eyal Neistein$^1$, Dan Maoz$^1$, Hans-Walter Rix$^{2,3}$, and John L. Tonry$^4$}

\noindent$^1$ School of Physics \& Astronomy and Wise Observatory, Tel-Aviv University,\\
\indent  Tel-Aviv 69978, Israel\\
$^2$ Steward Observatory, University of Arizona, Tucson, AZ 85726\\
$^3$ Alfred P. Sloan Fellow\\
$^4$ Institute for Astronomy, University of Hawaii, \\
\indent 2680 Woodlawn Dr., Honolulu, HI 96822 \\

\begin{abstract}

We present an $I$-band Tully-Fisher relation (TFR) for 18 nearby
S0 galaxies using kinematics derived from 
long slit spectroscopy of stellar absorption lines. Our estimates of 
the circular velocity, $V_c$, at 2-3 exponential disk scale lengths
account for line-of-sight projection
and for the stellar random motions through an
asymmetric drift correction. Uniform and accurate
distance calibration for all objects is available from surface brightness
fluctuation measurements of Tonry  et al. (1998).
Despite the care taken in estimating both $V_c$ and $M_I$,
the TFR shows an {\it intrinsic} scatter, $\sim 0.7$~mag in $M_I$,
or $0.15$ in ${\rm log}_{10}V_c$.
This result is surprising, as S0
galaxies appear to have both the simple kinematics of disk galaxies, and 
the simple stellar populations of early-type galaxies. 
Remarkably, in this sample of overall rotation-dominated 
galaxies, the central stellar velocity dispersion 
is a better predictor of the total $I$-band luminosity
(through the Fundamental Plane relations) than the circular speed  
at several exponential scale lengths.
Furthermore, the TFR zeropoint, or the mean stellar
$I$-band luminosity at a given $V_c$, differs by only $\sim 0.5$ mag
between
 our sample of S0s and Mathewson et al.'s (1992)
sample of late-type spirals, once both data sets are brought
onto a consistent distance scale. This offset is less than expected
if S0s are former spiral galaxies with prematurely
truncated star-formation ($\gtorder 4$~Gyrs ago).
\end{abstract}


\section{Introduction}
\label{int}

The Tully-Fisher relation (TFR) is a correlation between 
some measure of the maximal, or asymptotic, circular velocity of 
the disk and the integrated stellar luminosity of a 
galaxy. Since its discovery (Tully \& Fisher 1977) much 
effort has been invested in studying its manifestation at
various wavelengths, its dependence on different kinematic tracers,
and its differences among galaxy types.

Measures of the circular velocity have been derived from
either the 21cm H I line width
or from optical rotation curves (e.g., 
Mathewson, Ford, \& Buchhorn 1992; 
Raychaudhury et al. 1997;   Giovanelli
et al. 1997ab). The optical rotation curves in all these cases
were derived from H II emission lines. Courteau (1997) has
recently compared the TFRs based on H I widths and
optical rotation curves and finds basic agreement among them.
Integrated galaxy magnitudes were initially measured 
in the $B$-band and later in the $I$ and $H$ bands (see
Aaronson et al. 1979, 1986).  The slope (Aaronson \& Mould 1983),
the zero point, and the scatter of the TFR
depend on the band  (see Jacoby et al. 1992, for a summary; see also
Tully et al. 1998). 
The lowest scatter has been found in the $I$
band ($\sim 0.1$ mag, Bernstein et al. 1994).
Presumably, this is because the $B$ magnitude
is more influenced by dust extinction and short-lived stellar
populations, while the infrared magnitude is a more robust measure
 of the total stellar mass of the galaxy.

While the TFR serves as a fundamental tool for measuring
extragalactic distances, the physical
mechanism behind its existence is also of great interest.
Possible explanations for a well defined TFR are emerging 
(Aaronson et al. 1979; Schechter 1980; Eisenstein \& Loeb 1996;
Dalcanton, Spergel, \& Summers 1997; Mo, Mao, \& White 1998;
Elizondo et al. 1998; Heavens \& Jimenez 1999).
Self-regulated star-formation,
cosmologically-determined initial angular momentum distributions,
and adiabatic baryon infall all seem to play important roles.
Alternatively, Milgrom (1983; 1989) has advocated that 
his Modified Newtonian Dynamics (MOND), designed to explain
the rotation curves of galaxies without resorting to dark matter,
also naturally predicts a TFR. In the MOND picture, 
the instrinsic scatter in the TFR for a given galaxy population 
simply reflects the spread of mass-to-light ratios ($M/L$)
in the population.

Most observational efforts 
have focussed on the TFR for late-type spiral
galaxies, one extreme of the Hubble sequence.
Rubin et al. (1985) studied the TFR for Sa, Sb and Sc galaxies,
claiming a zero-point offset between Hubble types Sa and Sc  that corresponds
to $1.5$ magnitudes in I. However, Giovanelli et al. (1997b) found
an offset of only $\sim 0.3$ mag between these Hubble types,
and  Aaronson \& Mould (1983), Pierce \& Tully (1988), and Bernstein
et al. (1994) did not find a type dependence in their TFRs.
None of these authors derived a TFR for the next
Hubble type, S0s,
because it is difficult to measure their rotation curves using
H I or H II emission lines. Although $27\%$ of the S0s in Roberts et al. (1991) 
have H~I gas detected, in many of them the gas shows unusual characteristics, such
as large velocity dispersions and counter-rotating components, and single-dish
measurements often cannot reveal that the gas is concentrated in the inner 
regions or in an outer ring (e.g. Van Driel \& Van Woerden 1991).

In this paper we explore an analogous relation for
these earliest-type disk galaxies.  S0 galaxies were
classified by Hubble as a transition class between spirals 
and ellipticals (see  van den Bergh 1997, for a recent review),
 and in the RSA catalog (Sandage \& Tammann 1981)
they comprise 11\% 
of bright galaxies. The formation histories of S0s
are not well understood and are likely to be heterogeneous.
Their overabundance in cluster environments (Dressler 1980; see, e.g.,
Hashimoto \& Oemler 1998, for an update)
has led to suggestions
that they are the products of disk-galaxy collisions and mergers
(Schweizer 1986), or of gas stripping in
later types (Gunn \& Gott 1972). Numerical simulations
of gas and stellar dynamics indeed suggest that the merger of two
gas-rich disk galaxies of unequal mass can produce an object resembling an S0
(Hernquist \& Mihos 1995; Bekki 1998a,b). In this picture, the merger
induces a flow of gas to the central parts of the product galaxy,
where the gas is almost completely transformed into stars during
an induced central starburst. The  simulated merger products resemble
actual S0 galaxies in that they are much less gas-rich than their progenitors, 
contain a thickened disk, and exhibit little, if any,  spiral structure.
Observationally, this scenario is not free of problems, e.g., 
the absence of two distinct populations of globular clusters
(old and young) in early-type galaxies (Kissler-Patig, Forbes, \& Minniti 1998).
As an alternative, Van den Bosch (1998) and Mao \& Mo (1998) have
proposed that S0s form a continuum with later types. In the context
of hierarchical galaxy formation models,
the bulge-to-disk ratio is a tracer of the formation redshift and/or
the initial angular momentum of the dark halo in which the galaxy
formed.

In their gross stuctural properties, 
S0s are similar to ellipticals and share
the same Fundamental
Plane relations (\eg Jorgensen, Franx, \& Kjaergard 1996). Since the central velocity dispersions
of S0s and ellipticals are considerably higher than the 
rotation velocities of spirals of a given luminosity (which usually rise
quickly to their assymptotic values), it appears that
the mass-to-light ratio in the inner regions of galaxies increases when going to earlier types.
The existence and parameters of a TFR for S0s could help us place them
relative to ellipticals and spirals, and give a better understanding of the physical
mechanism behind the TFR. From a practical viewpoint, a tight TFR for S0s
could improve the
distance estimate to many clusters, where S0s are the dominant population.

To our knowledge, there has been only one published effort to measure a TFR
in S0 galaxies, by Dressler \& Sandage (1983). They found no evidence
for any actual correlation between stellar luminosity
and the observed mean stellar rotation speed.
However, their rotation curves had very limited
radial extent and were not corrected for projection effects nor for the stellar 
velocity dispersions. Furthermore, approximate (Hubble flow) distances were used, and
 the integrated blue magnitudes were based on photographic plates.
An intrinsic TFR for S0s may have therefore been lost in the observational 
noise.

In this paper we attempt to measure an $I$-band TFR in a sample of S0
galaxies. Rotation curves are obtained from 
major-axis long-slit optical absorption-line spectra.
In \S 2 we describe our sample and observations. In \S 3 we
describe the spectroscopic and photometric reduction and analysis,
present rotation curves, and derive the asymptotic circular velocities
of the galaxies. In \S4 we derive the TFR relation and discuss
its implications. Our conclusions are summarized in \S5.

\section{Sample and Observations}

\subsection{Sample Selection}

Because of the paucity of gas in S0s, the circular velocities
must be estimated from stellar absorption-line
kinematics, which requires fairly  high signal-to-noise (S/N) ratios.
We therefore first chose the brightest-possible sample of galaxies.
Second, to explore or to establish any TFR we need accurate and
independent distance estimates to our sample galaxies.
Since the brightest S0s are nearby, Hubble distances,
even when corrected for peculiar velocities using a large-scale
flow model, are unreliable. We therefore chose only S0s
whose distances have been measured by Tonry et al. (1998) using
the surface brightness fluctuation (SBF) method (Tonry \& Schneider 1988; 
Tonry et al. 1997;
Blakeslee et al. 1998).

Specifically, our sample criteria were as follows:
a) The galaxy is in the Tonry et al. (1998) sample; 
b) Heliocentric radial velocity $<2000$~km s$^{-1}$;
c) Declination $>-20^{\circ}$;
d) RSA classification S0/E, S0, SB0, S0/Sa, SB0/SBa or S0pec;
e) RC3 (de Vaucouleurs et al. 1991) $B$ magnitude $<12.6$.

These criteria lead to an initial sample of nearly forty galaxies,
of which we observed a sub-sample of 20, devoid of morphological 
peculiarities, and with inclinations $i\sim35^{\circ}-60^{\circ}$.
In this inclination range both the corrections for $\sin{i}$ and
for the line-of-sight integration through the disk (see \S 4)
are small.  Two of the galaxies, NGC 4406 and NGC 4472, were
subsequently excluded from the sample because they showed
little or no rotation, with their kinematics dominated
by random motions. These galaxies are perhaps more suitably labeled
as elliptical galaxies, which do not have a major disk component.

For three of the 18 galaxies in our final sample 
(NGC 936, NGC 3115, 
and NGC 7332) the stellar
kinematics have been studied by other authors,
and only photometric data were needed. For
NGC 3115 we used kinematic data from Capaccioli et al. (1993) and
Illingworth \& Schechter (1982). A deep study of NGC 7332 by Fisher,
Illingworth, \& Franx (1994) provided the kinematics for this galaxy.
Rotation curves for NGC 936 were taken from Kent (1987)
and Kormendy (1983; 1984). 
Table 1 lists the objects,
their parameters, and our sources of data.

\subsection{Observations}

Cousins $I$ band photometry of the sample galaxies was obtained at the Wise
Observatory 1m telescope using a Tektronix $1024\times1024$-pixel back-illuminated
 CCD with a scale of 0.696$\pm$0.002 arcsec pixel$^{-1}$. For each
galaxy we took $1-3$ exposures of 300 s each. Most of the images
were obtained on 1995 December 29, and a few were taken on 1996 December
15 and on 1997 February 17. All nights had photometric conditions and
photometric standard stars from Landolt (1992) were observed throughout each
night, and used to translate counts to $I$-band magnitudes.

The spectroscopic observations were also
obtained at the Wise Observatory 1m telescope.  We used the Faint
Object Spectrograph Camera (Kaspi et al.  1996) coupled to the above CCD.
A
2$^{\prime\prime}$-wide slit and a 600 line/mm grism, gave a
dispersion of $3.68$ \AA\ pixel$^{-1}$ in the 4000--7263 \AA\ range,
corresponding to a resolution of $\sim 300$ km s$^{-1}$.  The 
angular sampling was 2.08 arcsec pixel$^{-1}$.
 The observations were made on 1995, October 25--26, November 29,
and December 15--16, and 1996, March 14--16 and April 14--15. 
On each night we also obtained
spectra of bright stars, mostly K-giants, to serve as templates for
modeling the galaxy spectrum. Observations typically consisted of two
consecutive major-axis exposures for each galaxy.
Total integration times
varied from 1 hr for the brightest galaxies to 4 hrs for the
faintest. He-Ar lamp exposures, for wavelength calibration, and quartz lamp
exposures, for flat fielding, were taken between consecutive galaxy
exposures.
One spectrum of NGC 5866 was obtained at
Kitt Peak National Observatory (KPNO) using the
4m telescope on 1994 March 7, with the RC spectrograph,
a 1200 line mm$^{-1}$ grating and an exposure time of 30 min.

\section{Data Reduction and Analysis}

\subsection{Photometry}

The $I$-band images were reduced using standard IRAF\footnote{{IRAF
(Image Reduction and Analysis Facility) is distributed by the National
Optical Astronomy Observatories, which are operated by AURA, Inc.,
under cooperative agreement with the National Science Foundation.}} routines. Images
were bias subtracted, and flat-field corrected using twilight sky
exposures. Foreground stars were found and removed by examining each image
and replacing the affected area with
an interpolated two-dimensional surface, using the {\it Imedit} task.

In order to measure the ellipticity of each galaxy and the scale length
of its disk we used the {\it Ellipse} task.  
The semi-major axis lengths of the fitted elliptical 
isophotes was increased in
increments of 5\%
 until the change in the intensity between two
successive ellipses was negligible (except for two cases where a bright
star near the galaxy prevented extracting additional isophotes). The
task outputs for each ellipse the semi-major axis length, the mean
isophotal intensity, the ellipticity, and the position angle.
The {\it Elapert} task was then used to approximate each ellipse with a polygon and
the counts within each polygon were measured with the {\it Polyphot} task.
The projected disk ellipticity was taken
to be the ellipticity of the last well-fitted
ellipse. The disk scale length was found by $\chi^2$ minimization,
allowing the central surface-brightness, disk scale length, and sky
level to vary.
The parameters of the exponential disk fit and their 
uncertainties were used to extrapolate the counts from
the last measured radius to infinity, resulting in a ``total''
$I$-band magnitude and its error. Tonry et al.'s (1998) distances
to the galaxies were used to derive the absolute magnitudes, $M_I$.
These parameters are listed in Table 1.

\subsection{Spectroscopy}
The long slit spectra were also reduced using standard IRAF
routines. Each two-dimensional spectrum was bias subtracted.
Variations in slit illumination were removed by dividing 
each image by an illumination image derived from 
a spectrum of the twilight sky. Pixel-to-pixel sensitivity 
variations were removed by division by a quartz 
lamp spectrum taken after every galaxy exposure. 
The quartz spectrum was first normalized by a 6th-order 
polynomial fit to its low frequency
 structure in the dispersion direction.
 Cosmic ray events were removed with the IRAF tasks 
{\it Ccrej}, {\it Cosmicrays} and {\it Imedit}.
He-Ar arc-lamp spectra with about 40 lines were used to
rectify all science frames to uniform sampling
in slit position and $\log{\lambda}$, where $\lambda$ is the
wavelength, in the two cardinal
directions. The resulting accuracy of the wavelength calibration
is $\sim 15$~km s$^{-1}$.
The sky background was removed by interpolating
along the two ends of the slit, where the sky dominates.
Template star spectra were reduced in the same fashion,
and subsequently extracted from the frames to yield
one-dimensional spectra.

The line-of-sight velocities $V_{obs}(R)$ and velocity dispersions $\sigma(R)$
as functions of the projected radius $R$
were extracted from the galaxy spectra, following 
Rix \& White (1992) and Rix et al. (1995).
The two dimensional spectrum was first rebinned
into a sequence of one-dimensional spectra of
approximately constant S/N and each of these 
spectra was then matched by a shifted and broadened linear
combination of templates, minimizing $\chi^2$.
This resulted in a kinematic profile that, at each radius,
is derived from an ``optimal'' template.

Figure 1 (top and middle panels) shows rotation and 
velocity dispersion curves, $V_{obs}(R)$ and $\sigma(R)$,
for the 15 galaxies we observed spectroscopically.
One of the 
galaxies, NGC 5866, was measured both at Wise Observatory and at KPNO
(see Fig. 1). Although the degradation in S/N when going to a small
telescope is obvious, the agreement is good and shows that, for the
present purpose the Wise Observatory spectra are  
of sufficient quality.

\subsection{Deriving Circular Velocities from $V(R)$ and $\sigma(R)$}

Determining the true circular velocity of a galaxy,
defined as $V_c(R)\equiv \sqrt{R\frac{\partial\Phi_{grav}}{\partial
R}}$, from stellar kinematics is somewhat model-dependent, even if 
rotation dominates (see, e.g., discussion by 
Illingworth \& Schechter 1982; Binney \& Tremaine 1987;
Raychaudhury et al. 1997). 
We derive the circular velocity in several steps.
When several rotation curves were available for a single
galaxy (see \S 3.2) we computed the asymptotic velocity and
velocity dispersion in each curve separately and subsequently used the means.

To obtain the mean stellar rotation
velocity, $V_\phi$, in the plane of the disk, we deproject the observed
velocity, using the observed
disk ellipticity and assuming an edge-on disk axis ratio $q_0=0.22$ 
(de~Vaucouleurs et al. 1991):
\begin{eqnarray*}
 V_{\phi}(R)=\frac{ V_{obs}(R)}{ \sin(i)}=V_{obs}\times{\sqrt{\frac{
 1-q_{0}^{2}}{ 2e-e^{2}}}} \mbox{  ,}
\end{eqnarray*}
where $i$ is the inclination,
$e$ is the ellipticity, $V_{obs}$ is the observed radial velocity,
and $V_{\phi}$ is the azimuthal speed.
Galaxies with ellipticities greater than 0.57
($i>67^{\circ}$) were deemed to be edge-on, and no attempt at 
the above inclination correction was made.

However, in highly-inclined galaxies the
line-of-sight integration through the disk will reduce the
observed mean velocity relative to the actual velocity 
$V_\phi(R)$ at the tangent point.
We constructed a simple model of an exponential disk with 
a vertical scale height of $0.2R_{exp}$,
to calculate $V_{obs}/V_\phi(R)$.
For the edge-on case, an approximate analytic expression 
for $f\equiv V_{obs}/V_\phi(R)$ can be found, which is
is shown in Figure 2.
The same effect will lead to an overestimate of the
azimuthal velocity dispersion.
The two corrections for edge-on disks are:
\begin{eqnarray*} 
V_{\phi}(R)= \frac{
V_{obs}(R)}{f(\frac{ R}{ R_{exp}})} \mbox{  ,} 
\end{eqnarray*} 
\begin{eqnarray*}
\mbox{and  } \sigma_{\phi}^{2}=\sigma_{obs}^{2}-\frac{1}{2}(V_{\phi}-V_{obs})^{2} \mbox{  ,} \\
\end{eqnarray*}
\begin{eqnarray*}
\mbox{with  } f(x)=\frac{ \exp(-x)}{
-0.5772-\ln(x)+x-\frac{ x^{2}}{ 2\times2!} +\frac{ x^{3}}{
3\times3!}-...}-x , 
\end{eqnarray*} 
where $\sigma_{\phi}$ is the corrected velocity dispersion, 
and $\sigma_{obs}$ is the observed velocity dispersion.
Note that our uncertainties in how close to edge-on these galaxies
actually are, lead to an error of only 
$\Delta {\rm log}_{10}(V_{\phi})\approx 0.025$, assuming random
inclinations between $i=90^\circ$ and $i=70^\circ$.
For inclinations less than $70^{\circ}$, the correction is
 $<4\%$, and we neglect it.

Most importantly, however, $V_c$ will differ from the
directly observable quantities by the ``asymmetric drift''
correction, which accounts for the non-circular orbits of the
stars, or, equivalently, their velocity dispersion.
The circular velocity $V_{c}$ is related to the gravitational
potential $\Phi$(R), in the galaxy plane by
\begin{eqnarray*}
V_{c}^{2}(R)=R~\Bigl [\frac{ \partial \Phi(R)}{ \partial R}\Bigr ].
\end{eqnarray*}

To obtain the circular velocity (i.e. the velocity of a ``cold'' gas in
the disk) we follow Binney \& Tremaine (1987), eqn. 4-33:
\begin{eqnarray*}
V_{c}^{2}=\overline{V_{\phi}^{2}}+\sigma_{\phi}^{2}-\sigma_{r}^{2}
-\frac{ R}{ \rho}\frac{ \partial(\rho\sigma_{R}^{2})}{
 \partial R}-R\frac{ \partial (\overline{V_{R}V_{z}})}{ \partial z},
\end{eqnarray*}
where $\rho(R)=\rho_{0}\exp(-\frac{ R}{ R_{exp}})$ is the
mass density, and the term  $\overline{V_{R}V_{z}}$ is usually
negligible (Binney and Tremaine, 1987). For a flat rotation curve,
$\sigma_{\phi}^2(r)/\sigma_r^2(r)=0.5$, which leads to

$$V_{c}^2=V_{\phi}^2+\sigma_\phi^2\Biggl [ 2\Biggl( \frac{R}{R_{exp}} - 
\frac{\partial {\rm ln}~\sigma_R^2}{\partial {\rm ln}~R}\Biggr
)-1\Biggr ].$$
For many of the sample galaxies 
$\frac{\partial {\rm ln}~\sigma_\phi^2}{\partial {\rm ln}~R}$, and
hence $\frac{\partial {\rm ln}~\sigma_R^2}{\partial {\rm ln}~R}$,
is small, and can be neglected, yielding:
\begin{eqnarray*}
V_{c}^{2}=V_{\phi}^{2}+\sigma_{\phi}^{2}\Bigl (2\frac{ R}{ R_{exp}}-1\Bigr ) .
\end{eqnarray*}

To obtain the corrected rotation curves, we first 
fit an exponential function to the observed dispersion profile
 $\sigma_{\phi}(R)$. We then use the fit value of  $\sigma_{\phi}(R)$
to apply the asymmetric drift correction to every measurement
of $V_{\phi}$ for which $V_{\phi}/\sigma_{\phi}>2.5$ (see below).
The final, corrected, curves are shown in the bottom panels of 
Figure 1. Finally,
to estimate the deprojected, asymptotic rotation speed (usually at 
$R\sim 3R_{exp}$), we average the last three points on either side
of the corrected rotation curves in Figure 1 (bottom panels). 
Points with errors $\ge 100$~km s$^{-1}$ were discarded.
The radius of the measured aymptotic velocity, $R$, 
was taken as the average radius of the points in the rotation curve
that we used, and the uncertainty in that radius is half the
distance between the inner point and the outer point that we
used to obtain the final velocity.

We list all the measured and corrected velocities in Table 1.
Three of the galaxies, NGC 2768, NGC 4382, and NGC 4649, have
relatively large velocity dispersions even in their outer
parts, such that $\sigma_{\phi}\gtorder V_{\phi}/2.5$. Under
such circumstances, the approximations and systematics involved
in the asymmetric drift correction may lead to an unacceptably large 
error in the inferred $V_c$ and we mark the measurements of these galaxies 
as uncertain in 
the subsequent discussion.

\section{Results}

With the information assembled in Table 1 we can explore the two
questions posed initially: {\it a)} To what extent do S0s follow a TFR,
i.e., how well are $M_I$ and $V_c$ correlated? {\it b)} What is the
mean stellar luminosity for S0s at a given circular velocity, and how
does it compare to the luminosity of later-type disk galaxies?

Figure 3 shows $M_I$ {\it vs.} $V_c$ for the sample galaxies. The
errorbars in $M_I$ include photometric errors and distance
uncertainties, and the errors in $V_c$ include propagation of all the
uncertainties involved in the calculation of the final circular
velocity.  The data points with dotted errorbars represent the three
galaxies for which the asymmetric drift corrections were uncertain due
to their relatively large velocity dispersions (see above).  The dashed
line shows the $I$-band TFR for late type spiral galaxies, as derived
from the Mathewson et al. (1992) data by Courteau and Rix (1998) and
adjusted to the same distance scale ($H_0=80$~km s$^{-1}$ Mpc$^{-1}$)
as that implied by the SBF method for these galaxies (Tonry et al.
1998).

To estimate the best fit and the intrinsic scatter in the TFR,
we proceeded as follows (see also Rix et al. 1997.)
We assumed a relation of the form
$$
    M_I(\log{V_c}) = M_I( 2.3 ) - \alpha (\log{V_c}-2.3 ),
$$
 where the fit's pivoting point is 200 km s$^{-1}$, i.e., $\log{V_c}=2.3$.
Further, we assumed that the relation has an intrinsic Gaussian scatter in $M_I$ (at a given 
 $\log{V_c}$) of $\sigma$ magnitudes.
 For each parameter set $\Bigl [~M_I(2.3), \alpha, \sigma~\Bigr]$ this defines a model
 probability distribution, $P_{model}$ in the ($M_I,\log{V_c}$) plane.
 Each data point $i$, with its uncertainties in $V_c$ and $M_I$, also constitutes
 a probability distribution in the same parameter plane, $P_i(M_I,\log{V_c})$.
 The overall probability of a parameter set $\Bigl [~M_I(2.3), \alpha, \sigma~\Bigr]$,
 given the data, can be calculated as:
$$
 P(M_I (2.3),\alpha,\sigma)=\sum_{i}\int ~\Bigl (
   P_{model}\times P_{i}\Bigr )~
   \,dM\,d\log{V_c},
$$
 which is a measure of the overlap between the data and the model
probability distributions for a given model.

It is apparent from the data (Fig.~3) that the slope is poorly
determined.  Therefore, we fit a relation assuming the spiral TFR slope
from Mathewson et al. (1992), $\alpha=7.5$. The best fit has a
zeropoint of $ M_I(2.3) = -21.36\pm0.15 $ mag and an intrinsic scatter
of $\sigma = 0.68\pm0.15$ mag.  The thick line in Fig. 3 is this best
fit relation, and the thin lines show the scatter.
 From the plot is is clear that the data indicate a steeper slope.
 Formally, $\alpha > 10.5$ (at 95\% confidence), with no well-defined
 upper bound.
 Note that if we have underestimated
 the (dominant) velocity errors by 30\%, the estimated intrinsic scatter
 in the relation will only decrease to $\approx 0.58$ magnitudes. 

Based on Figure 3, we can now answer the two question posed above:

\noindent$\bullet$~ Despite the care taken in deriving $V_c$ and
$M_I$, there is a great deal of intrinsic scatter in the TFR:
$0.68\pm 0.15$ mag.

\noindent$\bullet$~ At a given $V_c$, 
there is only a small ($0.5\pm 0.15$ mag) systematic offset in $M_I$,
between the S0s and the
Sc galaxies from Mathewson et al. This offset is much smaller than the
1.5 magnitudes (in $I$) between Sa's and Sc's, claimed by Rubin et al. (1985),
and adds to the other evidence (e.g., Pierce \& Tully 1988;
 Bernstein et al. 1994) that the
zero point of the I-band TFR is only weakly dependent on galaxy type.

The large scatter in Fig.~3 is particularly remarkable in light
of the well-behaved Fundamental Plane (FP)
 relation (\eg Jorgensen et al. 1996,
and references therein) for S0s in general, as well as
for this particular set of objects. Figure 4 shows the FP for
our sample, based on values for the effective radii, $R_e$,
as compiled in Bender, Burstein, \& Faber (1992) and Fisher (1997), and
central velocity dispersions, $\sigma_0$, estimated
both from our data and the literature, and listed in Table 1.  
For comparison with the existing FP literature, we reconstructed 
$I_{eff}$ from $M_I$ and $R_e$, assuming a de~Vaucouleur's law.
The median scatter
among the points is well below $0.1$ in either axis.

The important difference between Figures~3 and 4 is that
the FP in Figure~4 uses the {\it central} stellar dispersion 
as the kinematic parameter, while the TFR in Figure~3
involves $V_c$ at 2--3$R_{exp}$, characterizing the total mass within this
radius.
It is clear from this comparison that, at least for this sample,
the central stellar dispersion
is a much better predictor of the total stellar luminosity than 
the circular velocity at several disk exponential radii.

We have searched
for possible sources, either observational or intrinsic, for
 the large scatter we have found in the S0 TFR. 
  Fisher (1997) obtained stellar rotation curves and velocity dispersion profiles for 18 
S0 galaxies, 7 of which are in our sample. 
Although he presents his measurements only out to about one disk scale length, $R_{exp}$, 
while our rotation curves typically extend to $R/R_{exp}=2-4$, a meaningful comparison can
 be made, since, as seen in Fig. 1, the rotation curves usually flatten out already at
 small radii (10 to 25 arcsec). Our measured asymptotic line-of-sight velocities agree with
 Fisher's at the $\sim 10\%$ level. A similar level of agreement exists between his measurements
in the $B$-band and our measurements in $I$-band of the disk
scale lengths and ellipticities. 
Velocity dispersions in his data  are also generally
consistent with ours, except for two cases, NGC~4382 and NGC~5866, in which he measures
 twice the values we obtained. NGC 4382, however, was already excluded from our analysis
above because of its relatively low level of rotation, while for NGC 5866 we have both 
Wise Observatory data and high-quality
data from KPNO, which are consistent with each other. 
Simien \& Prugniel (1997), Bettoni \& Galletta (1997), Fried \& Illingworth (1994) and 
Seifert \& Scorza (1996) have each derived rotation curves for some of the galaxies
in our sample, and their results are in good agreement with ours. A mild exception
is NGC 2549, for which Simien \& Prugniel (1997) and Seifert \& Scorza (1996) obtain a
 maximum velocity of $150\pm30$~km s$^{-1}$ compared to our $113\pm13$~km s$^{-1}$.  

While our sample has the advantage of uniform SBF distance estimation, distance
errors could contribute to the TFR scatter as well.
The SBF method has an r.m.s scatter of less than 0.1 mag, but
 there are a number of distance discrepancies which could affect a small sample like ours.
Among the galaxies in our sample, 
Blakeslee et al. (1998) and Ciardullo et al. (1993) find differences of order of 0.3 mag
between SBF-based distance modulii
and distances based on planetary nebula luminosity functions for
NGC~3115, NGC~4382, and NGC~1023. However, it is difficult to see how this
could be a dominant source of scatter in the TFR without introducing
comparable scatter in the Fundamental Plane relation for our sample.

A second potential source of errors is in the corrections for inclination and assymetric
drift we have applied to our data. These corrections are sometimes at a level of
 $100\%$ (most are above $35\%$) and are based on noisy velocity dispersion measurements.
H~I observations for some of our galaxies exist, and can partially confirm the velocity 
corrections. Comparisons of H~I velocities 
 and corrected stellar velocities 
are not straightforward, since the gas component in S0s may sometimes be concentrated only
in the inner parts or in an outer ring, as a relic from a past accretion event.
Furthermore, there are different measures of 21 cm linewidth (e.g., at $50\%$ or $20\%$ of the peak).
Nevertheless, 
from Roberts et al. (1991), Huchtmeier et al. (1995), and Wardle \& Knapp (1986) we obtained H~I 
velocities for five galaxies in our sample, and find excellent agreement with our corrected
stellar velocities in four cases, the exception being NGC~1052, where there is a $\sim 2
\sigma$ discrepancy between the inclination-corrected H~I width of Roberts et al. (1991), 
288 km s$^{-1}$, and our final circular velocity of $190\pm 39$ km s$^{-1}$.
As an alternative method of calculating the assymetric drift correction, we attempted, instead
of the procedure decribed above, to apply the correction
directly to the outermost measurements of the velocities and dispersions, 
after averaging the outer three points. However, this had the effect of increasing
 the scatter in the TFR. This is a consequence of the large $R/R_{exp}$ values making the 
dispersion term in the assymetric drift correction, 
$\sigma_{\phi}^{2}\Bigl (2\frac{ R}{ R_{exp}}-1\Bigr )$, dominant 
compared to $V_{\phi}^{2}$. Modifications in the choice of $R$, or correcting
 $\sigma_{\phi}$ for inclination had little effect on the TFR scatter.

Next, we searched for intrinsic sources of TFR scatter, arising from a possible
dependence on additional parameters. We have checked for correlations among the
residuals in the best-fitting TFR relation and a variety of parameters.
We found no dependence of TFR residuals on disk ellipticity, as was found, e.g.,
in the late-type-galaxy TFR of Bernstein et al. (1994) and interpreted as the effect of 
extinction by dust.
The ratio $ \frac {V_{\phi}} {\sigma_{0}}$ can serve as a kinematic indicator of 
rotational vs. dispersive support in a given galaxy, and correlation of the TFR
residuals with it could indicate, e.g., that those galaxies with the least rotation
(and the largest assymetric drift corrections) are those contributing most to
the scatter. However, we found no significant correlation between the TFR residuals
and this ratio. Similarly the residuals are not correlated with $\frac{R_{exp}} {R_{e}}$,
a photometric measure of disk vs. bulge dominance.
We found that the parameter $ x=\frac {V_{\phi}} {\sigma_{0}}-\frac{R_{exp}} {R_{e}}$  
is marginally correlated with the TFR residuals, at a significance level of 93\%.
Although the physical significance of $x$ is unclear, 
applying this correction reduces the intrinsic TFR scatter by $\sim 0.2$ mag.

In view of these tests, we conclude that the large intrinsic TFR scatter of 0.7 mag that 
we find for S0s is most likely not the result of errors in observation and analysis. Likewise,
we have not found additional parameters that significantly lower the scatter.
For comparison, the
TFR in late type spirals usually has an intrinsic r.m.s scatter of $\sigma_{in}\sim 0.25$ mag 
(e.g. Giovanelli et al. 1997b), although a smaller scatter can occur in homogeneous, well-defined samples;
Bernstein et al. (1994) found an r.m.s. scatter of 0.23 mag, which, after correction for 
extinction based on ellipticities reduced to 0.1 mag. 

  From the physical viewpoint,
our result is in conflict with the idea
that most S0s were disk galaxies -- on their way to become
present day spirals -- whose star-forming career was cut
short by some mechanism, e.g. tidal stripping in a dense
environment (Gunn \& Gott 1972).
In that case, we would expect the S0s to have faded significantly
at constant $V_c$, exhibiting a larger TFR zeropoint offset.
Specifically, if S0s had had similar star formation histories to Sc's
(e.g., Kennicutt et al. 1994) until a truncation,
say, $\gtorder 4$~Gyrs ago, we would expect an offset of $\gtorder 0.9$~magnitudes
in $I$ due to the fading of the stellar population,
based on Charlot \& Bruzual (1991) models.

Similarly, the absence of a tight S0 TFR argues against
a physical continuity of S0s with later-type spirals, as suggested
in the context of hierarchical structure formation models
(Van den Bosch 1998; Mao \& Mo 1998). Alternatively,
 S0s may be more closely related to ellipticals. Both may be 
the relics of non-cataclysmic mergers (Schweizer 1986). For individual sample members,
(\eg NGC 4649, NGC 4406, NGC 4472) this may be apparent from their individual structure,
but the present evidence is pointing towards this being true for a good fraction 
of the morphological class. Qualitatively,
the spread among S0s in time elapsed since the merger
and its ensuing gas-depleting starburst would produce the TFR scatter,
while, on average, the larger concentration of stars may 
compensate for the fading of the stellar population, and give
a mean luminosity comparable to that of late-type galaxies,
for a given halo mass. A quantitative examination of the TFR
resulting in this scenario is, however, needed.

\section{Conclusions}

We have constructed a TFR for nearby S0 galaxies,
deriving corrected circular velocities from stellar velocities,
and using high-quality distance estimates (Tonry et al. 1998)
based on surface brightness fluctuations. Despite the care taken, the relation
between $M_I$ and $V_c$ exhibits  $\sim 0.7$~magnitudes of scatter.
 As an illustration, NGC~2787 and NGC~4753 both have similar
 circular velocities of  230~km s$^{-1}$, but their
luminosities differ by over 3 mag.
The reason for this large scatter is not clear. Perhaps it indicates that 
the S0 morphological class truly represents a ``mixed bag", with
a wide range of galaxy formation channels feeding into it.
The central stellar velocity dispersion is a much
better predictor of the total stellar luminosity than $V_c$
at several exponential radii.

Similarly, the fact that on average S0's and Sc's of the same $V_c$ have
such similar luminosities is a puzzle. S0s 
have older,
and hence dimmer, stellar populations,
which should lead to
a TFR zero-point offset.
The absence of such an
offset could be explained if S0s have a considerably higher
fraction of their total mass in stars than Sc's. This perhaps
would be expected in the merger-formation scenario if, in fact,
such events are very efficient at converting the available gas
into stars. 

Observationally, it is desirable to reconfirm our result 
on a larger sample with higher
S/N measurements at larger radii, where presumably the kinematic 
corrections will be smaller. An independent test, which is insensitive
to errors in the distance estimate, is to measure the TFR for
S0s in a galaxy cluster. Analysis of such a measurement for the
Coma cluster is underway (Hinz, Rix, \& Bernstein 1999).

\acknowledgements
We thank Rachel Somerville for useful discussions, and the
referee, Brent Tully, for helpful comments.
This work was supported by the US-Israel Binational Science Foundation
Grant 94-00300, and by the Alfred P. Sloan Foundation (HWR).

\begin{figure}
\epsscale{1.1}
\plotone{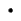}
\end{figure}

\begin{figure}
\epsscale{1.1}
\plotone{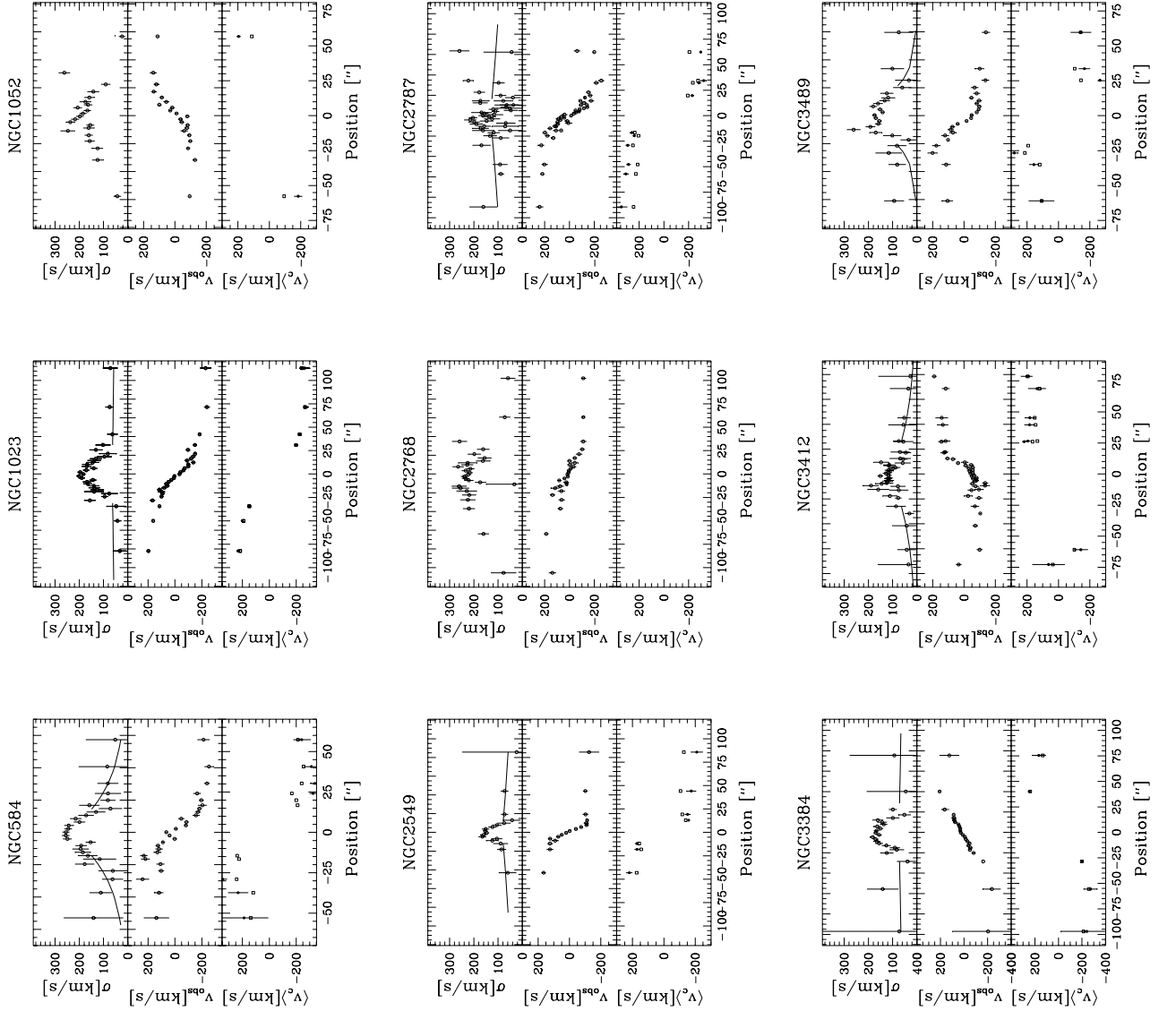}
\caption{Rotation and 
velocity dispersion curves, $V(R)$ and $\sigma(R)$,
for the 15 galaxies observed spectroscopically.
Top panels show the observed velocity dispersion profiles, $\sigma_{\phi}(R)$.
Middle panels show the observed rotation curves $V_{obs}(R)$.
Bottom panels show the observed rotation curves (empty symbols)
and the circular velocity (filled symbols), 
$V_c(R)$, after correction for inclination
or integration along the line of sight and asymmetric drift correction.}
\end{figure}
            
\begin{figure}
\epsscale{1.1}
\plotone{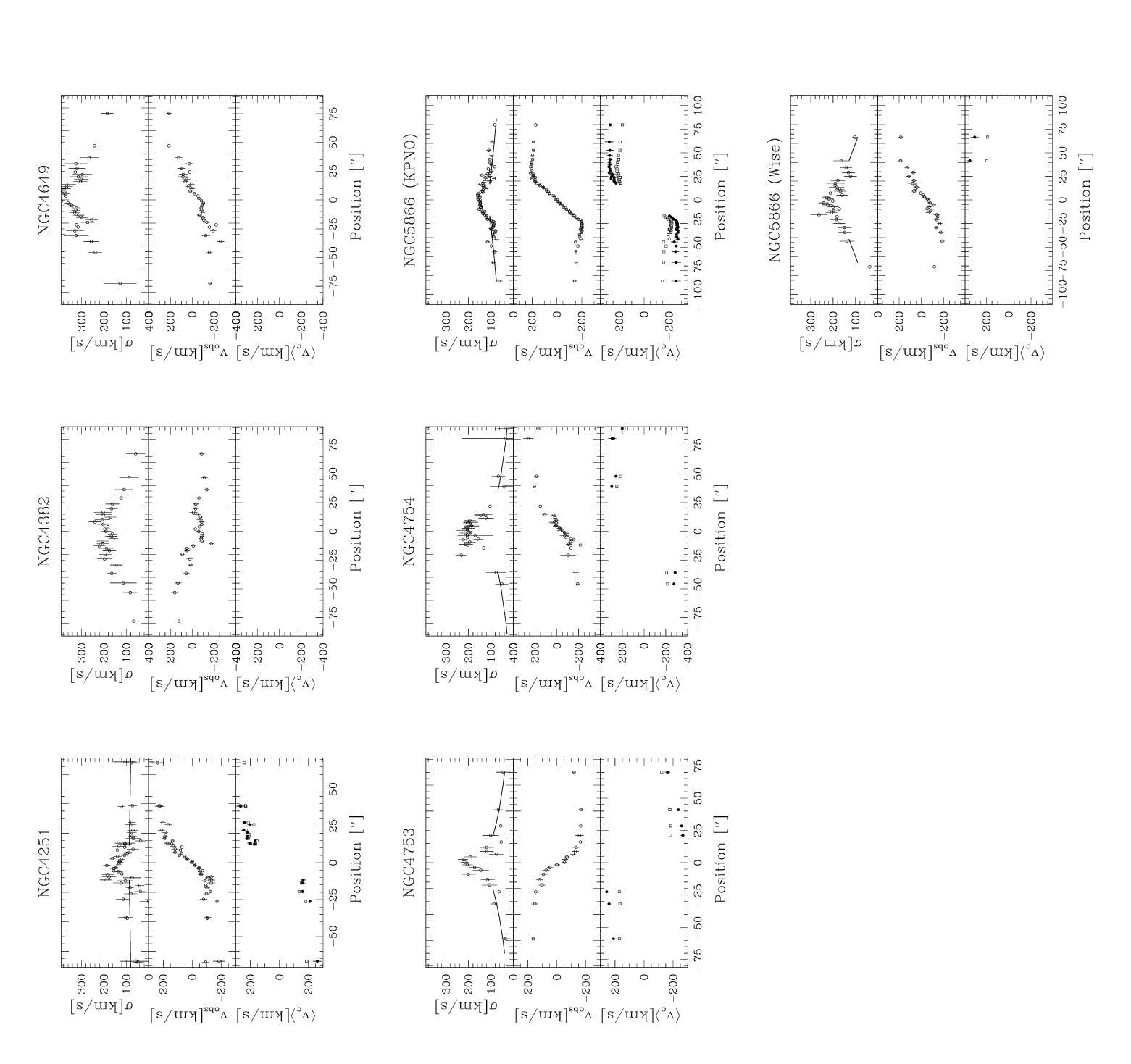}
\end{figure}
            
\begin{figure}
\epsscale{0.8}
\plotone{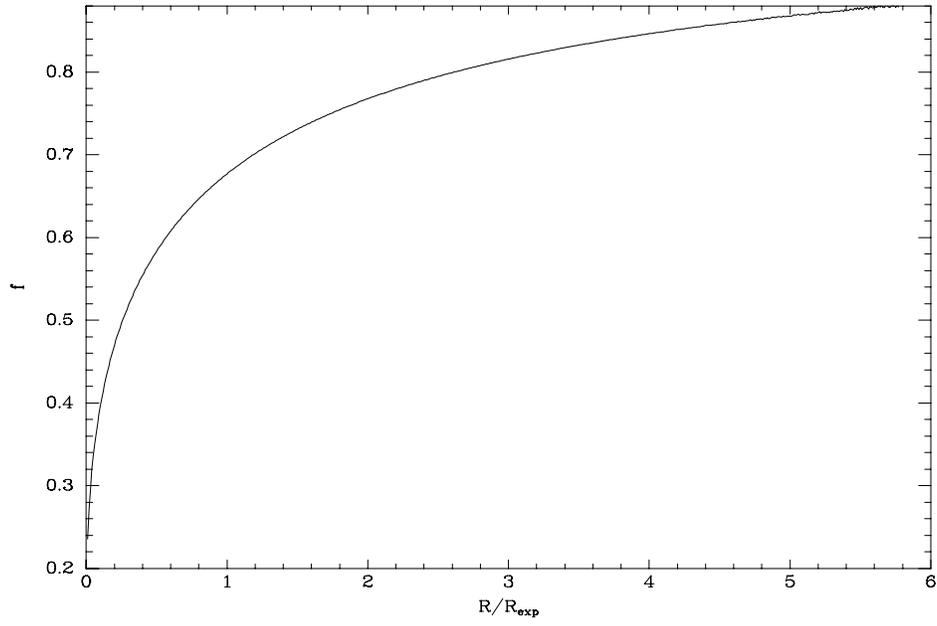}
\caption{The correction function $f(R/R_{\exp})$,
which we apply to the observed velocity of
nearly  edge-on galaxies to account for integration along the
line of sight.
}           
\end{figure}
            
\begin{figure}
\epsscale{1.0}
\plotone{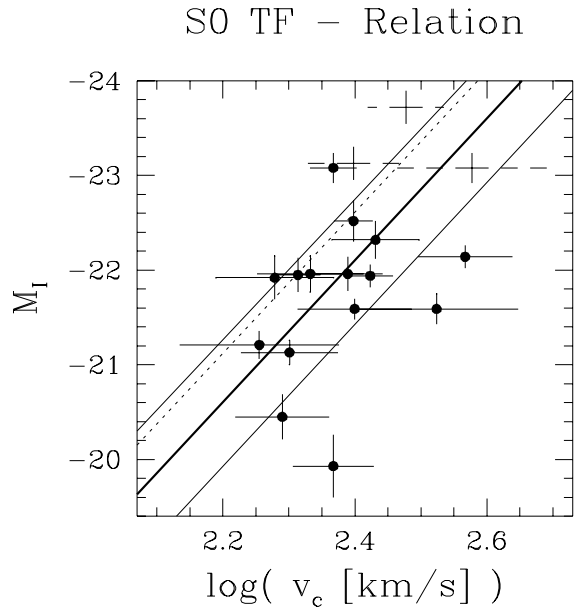}
\caption{$M_I$ {\it vs.} $V_c$ for our sample galaxies.
The data with dotted errorbars indicate the objects which 
have relatively large velocity dispersions even in their outer parts,
leading to asymmetric drift corrections $\ge 25\%$. 
The dashed line
shows the $I$-band TFR for late-type spiral galaxies, as derived from
the Mathewson et al. (1992) data by Courteau and Rix (1998) and adjusted
to the same distance scale ($H_0=80$~km s$^{-1}$ Mpc$^{-1}$) as that 
implied by the SBF method for these galaxies (Tonry et al. 1998).
The thick solid line shows the best fit relation, when constrained
to have the same slope as the late-types, and the thin lines mark
the intrsinsic scatter.
}           
\end{figure}
            
\begin{figure}
\epsscale{1.0}
\plotone{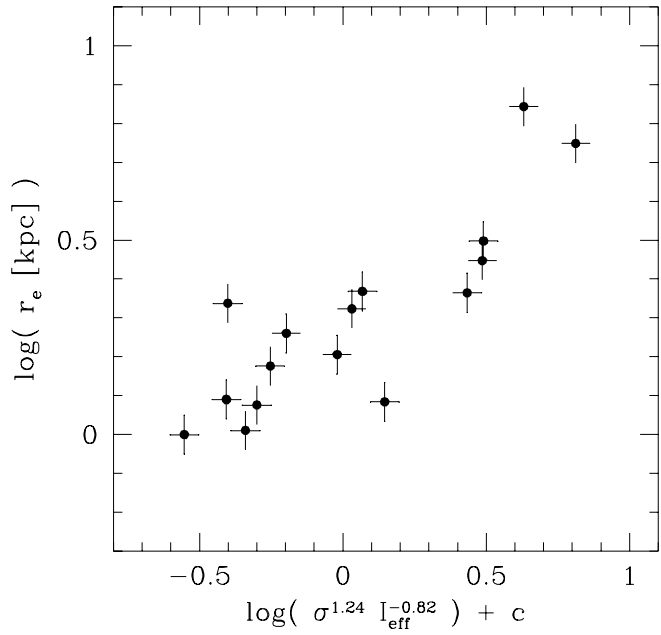}
\caption{Fundamental Plane relation for
our sample, based on values for the effective radii, $R_e$, and
and the central velocity dispersions, $\sigma_0$, taken both
from our data and from the literature. The median scatter
among the points is well below $0.1$ in either axis.}
\end{figure}

\end{document}